\documentclass[11pt,twoside]{article}
\usepackage{asp2010}
\usepackage{graphicx}
\usepackage{natbib}

\resetcounters

\begin{document}

\title{High Precision Simulations of the Evolution of a Super Star Cluster
Around a Massive Black Hole}
\author{R.~Capuzzo-Dolcetta, M.~Spera
\affil{Dep. of Physics, ``Sapienza'', Universit\'a di Roma, Piazzale Aldo Moro 2, 00185 Roma, Italy}}

\begin{abstract}
We present preliminary results of the application of a new sophisticated 
code which allows high precision integration of orbits of stars belonging 
to a dense stellar system moving in the vicinity of a massive black hole. 
This mimics the situation observed in the center of many galaxies, where a nuclear star 
cluster contains a massive black hole which, in the past, was, likely, an active engine 
of violent emission of radiation. The main scope of our work is the 
investigation of the relaxation of the super star cluster on a
sufficiently long time, together with the investigation of its 
feedback with the massive black hole.
\end{abstract}

\section{Introduction}
It is known, nowadays, that many galaxies harbor a supermassive black hole (SBH)
in their innermost regions.
In many cases, the massive black hole is surrounded by an  
ultra compact stellar system, the so called Nuclear Star Cluster (NSC). 
Only in the case of the Milky Way the NSC is resolved in its stellar 
components, thanks to the high sensitivity and spatial resolution of modern instruments.  
The formation of compact nuclei in galaxies has been studied, among others, by \citet{cap1}
which developed  the original \citet{trem} hypothesis of the M 31 nucleus formation 
via repeated infalls of globular clusters braked by dynamical 
friction, as alternative to the in-situ formation hypothesis. Nevertheless, 
the full $N$-body simulations performed so far (see, e.g., \citet{cdmio}, \citet{ant11})
have not been extended enough in time and with a sufficient resolution  
to reach fairly conclusive statements about it.

\section{Our new N-body code}

To the purposes discussed above, we developed a direct summation $N$-body code which 
includes a Hermite 6th order time integrator 
implemented by mean of a hierarchical block time-stepping. This code is written in C++, 
and uses MPI, OpenMP and CUDA to fully exploit the available computational power of modern hybrid computational clusters. 
Our first use of this new $N$-body code is to
studying the dynamical evolution of the Milky Way NSC over few relaxation 
times.

\begin{figure}[!ht]
\bigskip
\centering
\includegraphics[scale=0.78]{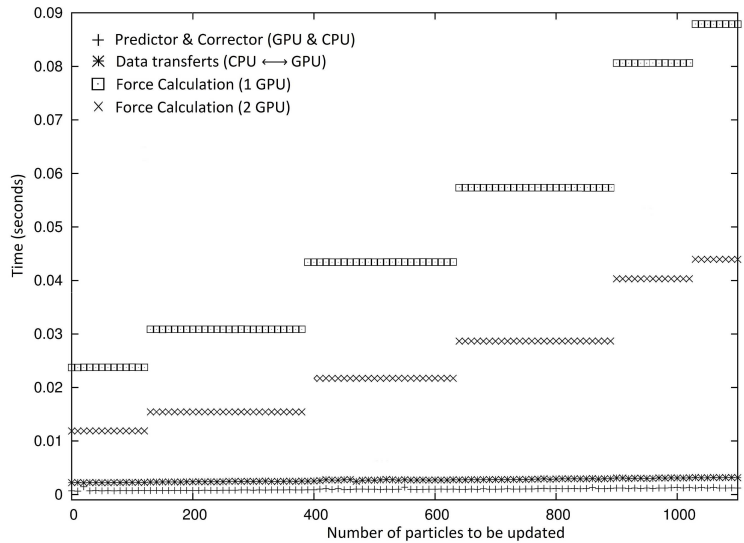}
\caption{Time spent in the execution of different parts of our code, in function of the number of particles 
to be time-updated.}\label{fig:Times}
\end{figure}

In modeling the NSC, we take as initial conditions a mass M$_{NSC}=1.32\times 10^7$~M$_\odot$ distributed over $N\simeq 270,000$ equal mass bodies,
and a central SBH of mass M$_{SBH}=4\times 10^6$~M$_\odot$. 
All the preliminary tests have been executed on our hybrid machine composed by 
2 NVIDIA Tesla C2050 GPUs in a host with 2 Intel Xeon X5650 esacore CPUs.
As we see in Fig. \ref{fig:Times}, the improvement of performance is crucial when the number of particles 
to be updated is small because, in this case, the time spent in transferring
data among CPUs and GPUs becomes comparable to that required by the force calculation 
(the GPU kernel). Moreover, we verified that the update 
frequency is much higher for small numbers of bodies (as expected given the presence of an SBH) 
and this constitutes the real bottleneck for such kind of simulations.

\section{Final remarks}
The simulation of the long term evolution of a Super Star Cluster around a massive 
black hole is a real computational challenge. At the light of the analysis presented here, 
we believe that a hybrid cluster of intermediate size fully exploiting 
the computational efficiency of our Hermite's 6th order code will allow 
to reach the goal of the description of the long term evolution of dense super star cluster around a massive black hole. Actually,
we estimate that a sufficiently extended (in time) simulation will require about 100 days 
of computation per node, as extrapolated from the preliminary benchmark runs done on our private machine.

\bibliography{CapuzzoDolcetta_R}

\end{document}